\documentclass[aps,amsmath,amssymb,amsfonts,twocolumn]{revtex4-2}
\usepackage{graphicx}
\usepackage[colorlinks=true,linkcolor=blue,citecolor=blue, urlcolor=blue]{hyperref}
\usepackage{blindtext}

\usepackage{lineno}

\usepackage{setspace}
\AtBeginDocument
{
	\addtolength\abovedisplayskip{0.20\baselineskip}%
	\addtolength\belowdisplayskip{0.20\baselineskip}%
}

\begin{document}

\title{Rapid state-recrossing kinetics in non-Markovian systems}

\author{Qingyuan Zhou}
\affiliation{Freie Universit\"at Berlin, Fachbereich Physik, 14195 Berlin, Germany}
\author{Roland R. Netz}
\affiliation{Freie Universit\"at Berlin, Fachbereich Physik, 14195 Berlin, Germany}
\author{Benjamin A. Dalton}
\affiliation{Freie Universit\"at Berlin, Fachbereich Physik, 14195 Berlin, Germany}

\begin{abstract}
The mean first-passage time (MFPT) is one standard measure for the reaction time in thermally activated barrier-crossing processes. While the relationship between MFPTs and phenomenological rate coefficients is known for systems that satisfy Markovian dynamics, it is not clear how to interpret MFPTs for experimental and simulation time-series data generated by non-Markovian systems. Here, we simulate a one-dimensional generalized Langevin equation (GLE) in a bistable potential and compare two related numerical methods for evaluating MFPTs: one that only incorporates information about first arrivals between subsequent states and is equivalent to calculating the waiting time, or dwell time, and one that incorporates information about all first-passages associated with a given barrier-crossing event and is therefore typically employed to enhance numerical sampling. In the Markovian limit, the two methods are equivalent. However, for significant memory times, the two methods suggest dramatically different reaction kinetics. By focusing on first-passage distributions, we systematically reveal the influence of memory-induced rapid state-recrossing on the MFPTs, which we compare to various other numerical or theoretical descriptions of reaction times. Overall, we demonstrate that it is necessary to consider full first-passage distributions, rather than just the mean barrier-crossing kinetics when analyzing non-Markovian time series data.
\end{abstract}



\maketitle

Thermally activated barrier-crossing processes are pervasive in chemical and biological systems. The association and disassociation of ion pairs \cite{Allen_1980, Trullas_1990}, molecular dihedral isomerization \cite{Rebertus_1979, Rosenberg_1980, Hochstrasser_1983, Troe_1992}, and the folding and unfolding of a protein \cite{Schuler_2008, Lindorff_2011, Chung_2013}, are all examples of systems that exhibit distinct configurational states, and which stochastically transition between states at a given rate. The dynamics of these systems are typically modeled as a one-dimension Markovian diffusion process over a free energy landscape, such that a state transition occurs when the system passages from one energy minimum to another, which it does by overcoming an energy barrier \cite{Kramers_1940,Hanggi_1990, Socci_1996, Best_2006}. While Markovian models are very useful and accurate in many contexts, it has become increasingly clear that non-Markovian effects, such as memory-dependent friction, cannot be neglected and must be accounted for to accurately predict many dynamic observables, such as the vibrational spectra of water \cite{Brunig_2022a}, ionic pair-reaction kinetics \cite{Brunig_2022d}, small-molecule dihedral isomerization kinetics \cite{Daldrop_2018, Dalton_arXiv_2023}, and fast protein-folding kinetics \cite{Ayaz_2021, Dalton_2023, Dalton_arXiv_2024}. While there exist multiple theories for describing reaction kinetics in non-Markovian systems \cite{Grote_1980, Pollak_1989, Kappler_2018, Kappler_2019}, it is still not entirely clear how the corresponding reaction rates, or times, predicted by the various theories should be extracted from experimental or simulation time series data.

\begin{figure}[t!]
\includegraphics[scale=1.02]{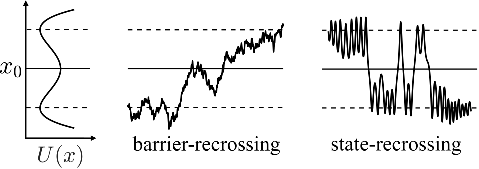}
\caption{Schematic comparison between barrier-recrossing and state-recrossing. A particle moves on a free energy surface $U(x)$ with a transition state located at $x_0$. Barrier-recrossing occurs when a particle makes multiple crossings of the transition state in a single excursion between the reactant and the product states. State-recrossing occurs when a particle reaches a product state and immediately, or soon thereafter, returns to the previous state. Here, the barrier-recrossing trajectory is generated by GLE simulation with $\tau_{\Gamma}/\tau_{\rm{D}} = 1.0{\times}10^{-3}$ and the state-recrossing trajectory with $\tau_{\Gamma}/\tau_{\rm{D}} = 1.0$.
 \label{Fig_0}}
\end{figure}

The various methods to evaluate barrier-crossing times from time-series data include evaluating either a barrier-escape time \cite{Skinner_1978, Talkner_1987b,Pollak_1989}, mean first-passage time (MFPT) \cite{Talkner_1987b, Talkner_1987a}, or the decaying time scale for some state-correlation function \cite{Chandler_1978, Roux_2022}. Escape times and MFPT require that an absorbing boundary is placed either at the barrier transition-state, or at some location in the product state, and one evaluates a mean time for a population of particles, initially located in the reactant state, to traverse the barrier and reach the absorbing boundary for the first time. The reactive flux formalism of Chandler \cite{Chandler_1978} relates the relaxation time scale of a population time-correlation function to the phenomenological population-decay time associated with first-order reaction kinetics. This relation, based on linear-response theory, explicitly accounts for a separation of time scales due to \textit{barrier-recrossing} effects, which, as we describe, are distinct from \textit{state-recrossing} effects (Fig.~\ref{Fig_0}). These methods are ideally suitable for evaluating overdamped kinetics in one dimension and the extension to multiple dimensions, or the low friction regime, typically requires some care \cite{Bagchi_2023}. However, these methods are not well characterized for non-Markovian systems. 

Here, we focus on the MFPT, evaluated for long, continuous simulations of one-dimensional, non-Markovian barrier-crossing dynamics. While the difficulties of treating MFPTs for non-Markovian processes have been discussed \cite{Talkner_1981, Hanggi_1983}, the direct numerical evaluation from extensive simulations has not been explored in detail. In particular, we compare two alternative numerical recipes for evaluating MFPTs. The first method calculates what is commonly referred to as the waiting time, or dwell time. Here, the time difference between first arrivals into neighboring states is considered and all crossings of a state minimum following a first arrival are neglected. The second method, however, incorporates information about every crossing of a state minimum such that many first-passage times are associated with a given transition. For Markovian systems, these two methods are in precise agreement. For systems that exhibit significant non-Markovian effects, however, the two methods present dramatically different mean reaction kinetics. This is due to the onset of rapid state-recrossing processes, which introduce fast non-Markovian modes into the first passage-time distributions. We further demonstrate that the transition-path time distributions are also significantly modified by the onset of an additional fast mode, which, in the limit of strong non-Markovianity, utterly dominates the mean transition-path time. Overall, we demonstrate that when analyzing experimental or simulation time series data for non-Markovian systems, simply considering the mean reaction kinetics can be misleading and that it is more revealing to consider the full kinetic distributions, which are indicative of the underlying non-Markovian mechanisms that determine the mean barrier-crossing kinetics.

\section*{Results}

We consider a particle moving in one dimension, coupled to a bath with a finite relaxation time, which we model using a generalized Langevin equation \cite{Zwanzig_1961, Mori_1965}:
\begin{equation}\label{GLE}
\begin{split}
m\ddot{x}(t) = -\int\limits_{0}^{t}\Gamma(t-t^{\prime})\dot{x}(t^{\prime})dt^{\prime} -\nabla U\big(x(t)\big)+  F_{\text{R}}(t).
\end{split}
\end{equation}
Here, $m$ is the mass of the particle, $\Gamma (t)$ is the friction memory kernel, $F_{\text{R}}(t)$ is the random force term satisfying the fluctuation-dissipation theorem $\langle F_{\text{R}}(t) F_{\text{R}}(t^{\prime}) \rangle =k_{\rm{B}}T\Gamma(t - t^{\prime})$, where $k_{\rm{B}}T$ is the thermal energy. $U(x)$ is the free energy profile and $\nabla \equiv \partial/\partial x$. The bath relaxation is introduced via the friction memory kernel for which we use a single-component decaying exponential $\Gamma (t) = (\gamma/\tau_{\Gamma})\text{exp}(-t/\tau_{\Gamma})$,  $\tau_{\Gamma}$ is the single bath relaxation time scale. To introduce a simple free energy barrier, we use a bi-stable potential where $U(x) = U_0[(x/L)^2-1]^2$. $U_0$ is the barrier height and $L$ is the distance between the minima. To simulate Eq.~\ref{GLE}, we use previously introduced Markovian embedding techniques (see \cite{Kappler_2018, Kappler_2019} and Appendix A for details). The other relevant time scales for this system are the inertia time $\tau_{\rm{m}} = m/\gamma$, where $\gamma = \int_0^{\infty} dt \Gamma (t)$ is the total friction acting on the particle, and the diffusion time $\tau_{\rm{D}} = \gamma L^2/k_{\rm{B}}T$, which we use to rescale time throughout.

\begin{figure}[b!]
\includegraphics[scale=0.98]{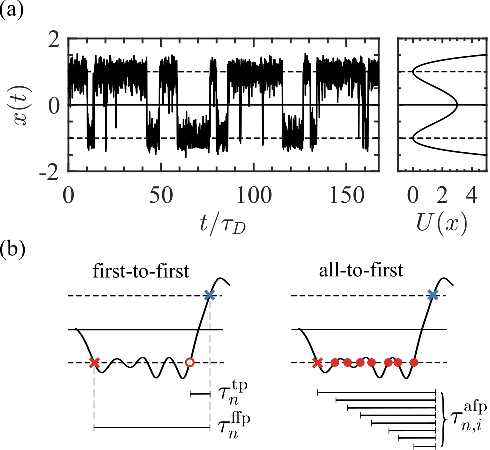}
\caption{(a) A typical trajectory segment for a one-dimensional GLE simulation (Eq.~\ref{GLE}) in a bistable well ($\tau_{\Gamma}/\tau_{\rm{D}} = 0.1$, $\tau_{\rm{m}}/\tau_{\rm{D}} = 0.001$, and $U_0 = 3\; k_{\rm{B}}T$). (b) Schematic illustration of the two passage-time definitions. The crosses show the initial entries into a new state and the filled circles show all subsequent crossings of the local minimum of that state. $\tau^{\rm{ffp}}_n$ is a single first-to-first passage time, which also includes the time to traverse the transition path $\tau^{\rm{tp}}_n$. $\tau^{\rm{afp}}_{n,i}$ is a sequence of all-to-first passage times corresponding to the single $\tau^{\rm{ffp}}_n$ such that $\tau^{\rm{afp}}_{n,1} = \tau^{\rm{ffp}}_n$. Here, $i=1,2,...,8$.
 \label{Fig_1}}
\end{figure}

In fig.~\ref{Fig_1}(a), we show a typical trajectory generated by simulating Eq.~\ref{GLE} with a moderate $\tau_{\Gamma}$. The particle diffuses under the influence of the bi-stable free energy profile and stochastically transitions between the two states by overcoming the free energy barrier, where here $U_0 = 3k_{\rm{B}}T$. To determine the mean barrier-crossing time, we distinguish between two MFPT approaches, which we schematically represent in Fig.~\ref{Fig_1}(b). The first-to-first passage times, which are also traditionally referred to as waiting times, or dwell times, only consider the time intervals between the initial entries into subsequent states. An individual first-to-first passage time $\tau_{n}^{\rm{ffp}}$, where $n$ in the index of the passage event, includes the corresponding transition-path time for that event $\tau_{n}^{\rm{tp}}$, which we also indicate in the schematic. The mean first-to-first passage time is given by $\tau_{\rm{ffp}} \equiv \langle \tau_{n}^{\rm{ffp}} \rangle_n$, where $\langle ... \rangle_n$ is the ensemble average over $n$, and hence the mean transition path time is $\tau_{\rm{tp}} \equiv \langle \tau_{n}^{\rm{tp}} \rangle_n$. As can be seen in Fig.~\ref{Fig_1}B, a single first-to-first passage event will spawn a sequence of all-to-first passage events as the particle oscillates around the local minimum. Therefore, for a given $\tau_{n}^{\rm{ffp}}$, there is an associated sequence $\tau_{n,i}^{\rm{afp}}$, where $\tau_{n,1}^{\rm{afp}} = \tau_{n}^{\rm{ffp}}$ and $\tau_{n,i}^{\rm{afp}} < \tau_{n}^{\rm{ffp}}$ for all $i>1$. The mean all-to-first passage time, which is given by $\tau_{\rm{afp}} \equiv \langle \tau_{n}^{\rm{afp}} \rangle_{n,i}$, where $\langle ... \rangle_{n,i}$ is the ensemble average over $n$ and $i$, is often used to achieve improved statistics when calculating MFPTs. However, as we show here, in the presence of memory, these times differ.

\begin{figure*}[t!]
\includegraphics[scale=1.0]{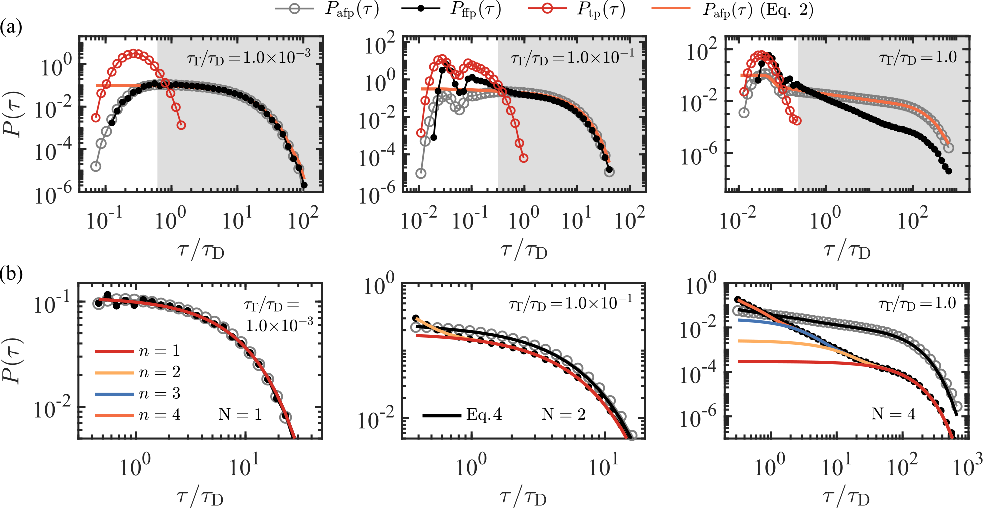}
\caption{Barrier-crossing and transition-path time distributions for systems with memory-dependent friction. (a) Full distributions for the all-to-first passage times $P_{\rm{afp}}(\tau)$, the first-to-first passage times $P_{\rm{ffp}}(\tau)$, and the transition path times $P_{\rm{tp}}(\tau)$. Distributions are accumulated from long equilibrium trajectories (see Fig.\ref{Fig_1}(a), total run time $1.25{\times}10^6 \tau_{\text{D}}$ per system) into histograms with exponentially-spaced bin widths. The orange curves show $P_{\rm{afp}}(\tau)$ predicted by mapping from simulation results for $P_{\rm{ffp}}(\tau)$ via Eq.~\ref{Dist_Rel_main}. $\tau_{\text{m}}/\tau_{\text{D}} = 1{\times}10^{-3}$ for all systems. (b) Replotted distributions from (a) (grey region) with an exponential series (Eq.~\ref{Exp_ffp}) fit to $P_{\rm{ffp}}(\tau)$, where $N$ is the minimum number of components required to fit a distribution. $N$ increases for increasing $\tau_{\Gamma}$. Exponential components are written here such that $\tau^{\text{exp}}_1>\tau^{\text{exp}}_2>\tau^{\text{exp}}_3>\tau^{\text{exp}}_4$. Black curves are for Eq.~\ref{Exp_afp}, with $\alpha_n$ and $\tau_n$ taken from fits of Eq.~\ref{Exp_ffp} to $P_{\rm{ffp}}(\tau)$.
 \label{Fig_2}}
\end{figure*}

The normalised distributions for the first-to-first and all-to-first passage times are given by $P_{\rm{ffp}}(\tau)$ and $P_{\rm{afp}}(\tau)$, respectively. In Fig.~\ref{Fig_2}(a), we show $P_{\rm{ffp}}(\tau)$ and $P_{\rm{afp}}(\tau)$ accumulated from long simulation trajectories for three representative memory times. The distributions are accumulated using exponentially increasing bin widths, which reveals details in both the long and short crossing-time regimes. Additionally, we also show the distributions for the transition-path times $P_{\rm{tp}}(\tau)$. Contributions from the transition-path times are clearly discernible in the short-time regimes of first-passage distributions, especially for the systems with longer memory times. For all simulations, we set $\tau_{\text{m}}/\tau_{\text{D}}=1{\times}10^{-3}$, such that inertial effects are negligible. Therefore, the system with the shortest memory time ($\tau_{\Gamma}/\tau_{\text{D}} = 1{\times}10^{-3}$) is representative of the overdamped, Markovian limit. In this case, $P_{\rm{ffp}}(\tau)=P_{\rm{afp}}(\tau)$, which, as we show below, indicates that the barrier-crossing kinetics for this system are described by single-component exponential distributions, which is a characteristic of overdamped Markovian systems. For significant memory times, which we take to be $\tau_{\Gamma}/\tau_{\text{D}} \sim 5{\times}10^{-2}$ (see below), $P_{\rm{ffp}}(\tau)$ and $P_{\rm{afp}}(\tau)$ diverge, and the divergence increases for increasing memory time. This divergence is a signature of non-Markovian state-recrossing dynamics.

In Appendix B, we show that $P_{\rm{ffp}}(\tau)$ and $P_{\rm{afp}}(\tau)$ are related by 
\begin{equation}\label{Dist_Rel_main}
P_{\rm{afp}}(\tau) =\frac{ \int\limits_{\tau}^{\infty}P_{\rm{ffp}}(\tau^{\prime})d\tau^{\prime}}{\int\limits_{0}^{\infty}P_{\rm{ffp}}(\tau^{\prime})\tau^{\prime}d\tau^{\prime}},
\end{equation}
This is confirmed in Fig.~\ref{Fig_2}(a), where the prediction of $P_{\rm{afp}}(\tau)$ from $P_{\rm{ffp}}(\tau)$ (orange lines) agrees well with the numerical results, breaking down only in the short-time regime where the distributions are dominated by transition-paths. While the prediction is trivial for the short and intermediate memory times, since the two distributions are equivalent, Eq.~\ref{Dist_Rel_main} maps accurately between $P_{\rm{ffp}}(\tau)$ and $P_{\rm{afp}}(\tau)$ for long memory time ($\tau_{\Gamma}/\tau_{\text{D}} = 1.0$), where the two distributions strongly diverge. The underlying grey regions indicate where the relationship between the two distributions is well described by Eq.~\ref{Dist_Rel_main}. In Fig.~\ref{Fig_2}(b), we re-plot $P_{\rm{ffp}}(\tau)$ and $P_{\rm{afp}}(\tau)$ corresponding to these regions, thereby excluding the transition-path-dominated regime. As such, the various distributions are well represented as exponential series. We fit $P_{\rm{ffp}}(\tau)$ such that 
\begin{equation}\label{Exp_ffp}
P_{\rm{ffp}}(\tau) = \frac{\sum_{n=1}^N \alpha_n\text{e}^{-\tau/\tau^{\text{exp}}_n}}{\sum_{n=1}^N \alpha_n\tau^{\text{exp}}_n},
\end{equation}
where $N$ is the number of components required to fit the data, which varies depending on $\tau_{\Gamma}$. The denominator is included to ensure that the distribution is normalized. From Eq.~\ref{Dist_Rel_main}, we obtain
\begin{equation}\label{Exp_afp}
P_{\text{afp}}(\tau)=\frac{\sum_{n=1}^N \alpha_n\tau^{\text{exp}}_n\text{e}^{-\tau/\tau^{\text{exp}}_n}}{\sum_{n=1}^N \alpha_n(\tau^{\text{exp}}_n)^2 },
\end{equation}
which is also normalised.

\begin{figure}[t!]
\includegraphics[scale=0.90]{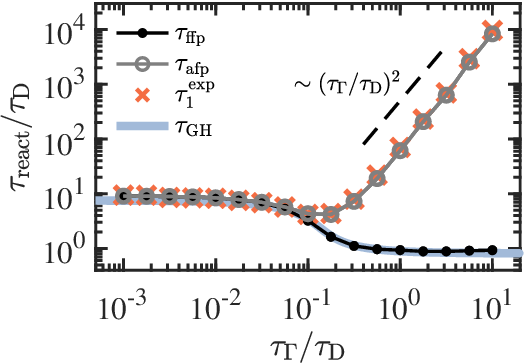}
\caption{Comparison of various definitions of barrier-crossing reaction times ($\tau_{\text{react}}$) for systems with memory-dependent friction. $\tau_{\rm{ffp}}$ and $\tau_{\rm{afp}}$ are the mean first-to-first and all-to-first passage times, respectively. $\tau^{\rm{exp}}_1$ is the fitting result for the longest-time-scale exponential components to the distributions $P_{\text{ffp}}(\tau)$ and $P_{\text{afp}}(\tau)$, as given in Eqs.~\ref{Exp_ffp} and \ref{Exp_afp}, and $\tau_{\rm{GH}}$ is the Grote-Hynes prediction for a single component exponential memory kernel. The black-	dashed line indicates quadratic scaling.
 \label{Fig_3}}
\end{figure}

In Fig.~\ref{Fig_2}(b), we show the individual components for an exponential series fit to each $P_{\rm{ffp}}(\tau)$. As expected, the system with the shortest memory time is well described by a single-component exponential ($N=1$), indicating that barrier crossing events are independent and uncorrelated. The mapping between $P_{\text{ffp}}(\tau)$ and $P_{\text{afp}}(\tau)$ is therefore trivial in this case, since the two distributions are equivalent. For intermediate $\tau_{\Gamma}/\tau_{\text{D}}=1.0{\times}10^{-1}$, we observe the onset of correlated transitions, which manifest as an additional fast exponential mode in the passage-time distributions ($N=2$). Upon surmounting the barrier and entering a new state, the particle may have completely thermalized, but the environment has not fully relaxed. In some instances, the relaxation process may cause the particle to rapidly recross the barrier, thereby increasing the probability of observing a fast-crossing event. Given that our simulations are overdamped by design, this enhancement of rapid state recrossing is purely due to memory-dependent friction effects. The excited fast mode is not visibly discernible in $P_{\text{afp}}(\tau)$ (Fig.~\ref{Fig_2}(b)). This is confirmed by the mapping prediction (Eq.~\ref{Exp_afp}, solid-black lines in Fig.~\ref{Fig_2}(b)). Here, the longest mode is amplified by a factor $\tau_1=4.1\tau_{\text{D}}$ compared to $\tau_2=0.14\tau_{\text{D}}$ for the fast mode. As the memory time is increased ($\tau_{\Gamma}/\tau_{\text{D}} =1.0$),we see the excitation of many faster modes in $P_{\text{ffp}}(\tau)$ and $P_{\text{afp}}(\tau)$. For $\tau_{\Gamma}/\tau_{\rm{D}} = 1.0$, we require $N=4$. The spread of time scales indicates that memory effects enhance not just immediate recrossing processes, but a range of short and intermediate state residencies, during which time the slowly relaxing environment can still induce recrossing events. Both $P_{\text{ffp}}(\tau)$ and $P_{\text{afp}}(\tau)$ reveal the presence of multiple exponential modes and the mapping between the two distributions via Eqs.~\ref{Dist_Rel_main}-\ref{Exp_afp} is ideal.

\begin{figure}[b!]
\includegraphics[scale=1.0]{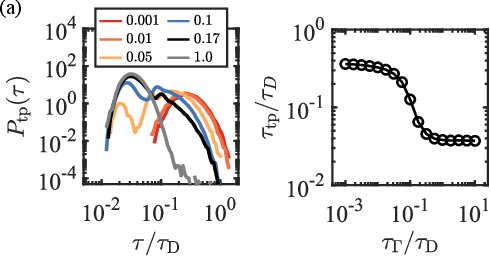}
\caption{
(a) Transition-path-time distributions over a range of memory times. The distributions reveal a transition between two modes of behavior, one dominated by overdamped Markovian barrier crossing and one dominated by non-Markovian recrossing where slow energy diffusion effects are strong. The distributions in the intermediate regime are discernibly bi-modal. The values in the legend are for $\tau_{\Gamma}/\tau_{\rm{D}}$. (b) The mean transition-path times $\tau_{\rm{tp}}$ for all memory times. The sigmoidal behavior shows the switching between the two characteristic modes in (a).
 \label{Fig_4}}
\end{figure}

By considering the MFPTs corresponding to each distribution, we appreciate how the choice of distribution can severely influence the interpretation of the system reaction kinetics. In Fig.~\ref{Fig_3}, we plot the mean first-to-first ($\tau_{\rm{ffp}}$) and mean all-to-first ($\tau_{\rm{afp}}$) passage times. $\tau_{\rm{afp}}$ exhibits three characteristic regimes, which have been discussed previously for a single-component exponential memory kernel \cite{Kappler_2018}. For small $\tau_{\Gamma}$, the system approaches the overdamped Markovian limit. For intermediate memory times, we observe a memory-induced accelerated barrier-crossing regime, and for long memory times, we observe a regime of memory-induced reaction slowdown, where $\tau_{\rm{afp}}$ increases quadratically with $\tau_{\Gamma}$. Interestingly, the time scale for the longest exponential mode ($\tau^{\rm{exp}}_1$), extracted by fitting each distribution with either Eq.~\ref{Exp_afp} or Eq.~\ref{Exp_ffp}, agrees very well with $\tau_{\rm{afp}}$. The linear contribution of $\tau^{\rm{exp}}_n$ in Eq.~\ref{Exp_afp} ensures that the distribution $P_{\text{afp}}(\tau)$ is dominated by slower modes. Therefore, $\tau_{\rm{afp}}$ emphasizes the uncorrelated contributions to the barrier-crossing kinetics, which stem from events where the particle has resided in the reactant state for long enough that memory of any previous transitions have subsided. $\tau_{\rm{ffp}}$, however, is strongly influenced by the non-Markovian recrossing processes at shorter time scales. In the overdamped, Markovian limit ($\tau_{\Gamma} \rightarrow 0$), $\tau_{\rm{ffp}}$ and $\tau_{\rm{afp}}$ are in exact agreement, which is clear from Fig.~\ref{Fig_2}, where $P_{\text{ffp}}(\tau)$ and $P_{\text{afp}}(\tau)$ for $\tau_{\Gamma}/\tau_{\text{D}} = 0.001$ are equal to $N=1$. However, $\tau_{\rm{ffp}}$ accelerates throughout the intermediate regime and then plateaus such that $\tau_{\rm{ffp}}(\tau_{\Gamma} \rightarrow \infty)<\tau_{\rm{ffp}}(\tau_{\Gamma} \rightarrow 0)$, hence revealing dramatically different behavior compared to $\tau_{\rm{afp}}$. One well-known theory for predicting barrier-crossing kinetics for systems with memory-dependent friction is the Grote-Hynes theory. Here, the traditional transition-state-theory prediction is corrected by a coefficient that explicitly depends on the curvature of the barrier top and the dominant relaxation time scale for a population of particles with a memory kernel $\Gamma(t)$ initially placed at the barrier top, which is an unstable state. Since the Grote-Hynes theory is derived from the perspective of a particle initiated from the transition state, and does not account for the particle history stemming from unsuccessful barrier crossing attempts, it can not capture the long time-scale kinetics associated with $\tau_1^{\text{exp}}$, which are shown to dominate $\tau_{\rm{afp}}$. We confirm this in Fig.~\ref{Fig_3}, where we show predictions by the Grote-Hynes theory $\tau_{\text{GH}}$. The results for $\tau_{\text{GH}}$ coincides with $\tau_{\rm{ffp}}$, indicating that Grote-Hynes theory is suitable for predicting waiting times, or dwell times, which are often measured in simulation and experiment. 


The signatures of non-Markovian recrossing also characterize the transition-path time distributions $P_{\rm{tp}}(\tau)$, which exhibits interesting bi-modal behavior (Fig.~\ref{Fig_4}(a)). In the short memory-time regime, we see a single, stationary mode, which is evidenced by the approximately constant mean value $\tau_{\text{tp}}$ for $\tau_{\Gamma}/\tau_{\text{D}} < 0.03$ (Fig.~\ref{Fig_4}(b)). Throughout the intermediate memory-time regime, we see the emergence of a second peak in the distribution for faster transitions. This faster transition mode represents the rapid rebound associated with the memory-induced recrossing. For increasing memory time, the recrossing transitions become the dominant mode of transition. This is clear from $P_{\rm{tp}}(\tau)$ for $\tau_{\Gamma}/\tau_{\text{D}} = 1.0$, which is almost purely comprised of such fast transitions. The sigmoid-like behavior of $\tau_{\text{tp}}$ in Fig.~\ref{Fig_4}(b) can therefore be interpreted as a transition between two modes of transition behavior, with the intermediate regime representing the shift in dominance. The same bi-modality emerges in the fast-crossing regime of $P_{\rm{ffp}}(\tau)$ (see the white underlaid region of Fig.~\ref{Fig_2}(a)), indicating that immediate recrossings overwhelm $P_{\rm{ffp}}(\tau)$ for long memory times. To quantify this, we note that the region of $P_{\rm{ffp}}(\tau)$ associated with immediate recrossing (white underlaid region in Fig.~\ref{Fig_2}(a)) contributes 93$\%$ of the total distribution weight for $\tau_{\Gamma}/\tau_{\text{D}} = 1.0$. For comparison, the same region of $P_{\rm{afp}}(\tau)$ contributes 5$\%$ to the total. Therefore, for significant memory times, the non-Markovian recrossing processes dominate both $\tau_{\text{ffp}}$ and $\tau_{\text{tp}}$. 

\section*{Conclusions}

In this article, we are concerned with the treatment of experimental or simulation time series data, and with the interpretation of kinetic information extracted from such data, when the system under observation exhibits significant non-Markovian effects. Various theoretical treatments exist for predicting barrier-crossing reaction times in the presence of time-dependent friction \cite{Grote_1980, Pollak_1989, Kappler_2018, Kappler_2019}. However, the precise agreement between theoretical rates and numerical techniques applied to non-Markovian data remains unclear. Here, by focusing on barrier-crossing distributions, rather than just the mean barrier-crossing kinetics, we have clarified the relationship between two standard numerical recipes for calculating MFPTs and we provide an exact theoretical relationship between the corresponding first-passage time distributions. While we specifically focus on non-Markovian processes in the overdamped limit, it should be noted that the relationship holds for underdamped Markovian systems with significant inertia, where state-recrossing also dominates the reaction kinetics \cite{Bagchi_2023}, since the same relationships between the first-to-first passage times $\tau^{\rm{ffp}}_{n}$ and the corresponding all-to-first passage times $\tau^{\rm{afp}}_{n,i}$ hold. Overall, we reveal that in the presence of significant state-recrossing effects, mean reaction times can emphasize either the rapid state-recrossing processes or the long-time-scale independent processes. Interestingly, the Grote-Hynes theory, which assumes initialization from the barrier top, predicts kinetic measurements that are dominated by state-recrossing. In all instances, the relationships between the various numerical definitions of barrier-crossing times are clearly understood from the perspective of the barrier-crossing distributions, which we suggest are more revealing than the mean barrier-crossing kinetics for systems that exhibit significant non-Markovianity.

\section*{ACKNOWLEDGMENTS}

We acknowledge support by the ERC Advanced Grant No. 835117 NoMaMemo and the Deutsche Forschungs- gemeinschaft (DFG) Grant No. SFB 1078.

\appendix

\renewcommand{\theequation}{A\arabic{equation}} 
\setcounter{equation}{0}

\section{Markovian embedding and the GLE}

To simulate Eq.~\ref{GLE}, we use previous Markovian embedding techniques \cite{Kappler_2018, Kappler_2019}. We note that we can rescale Eq.~\ref{GLE} for $\Gamma (t) = (\gamma/\tau_{\Gamma})\text{exp}(-t/\tau_{\Gamma})$ and generate a dimensionless form:
\begin{equation}\label{GLER}
\begin{split}
	\frac{\tau_{\rm{m}}}{\tau_{\rm{D}}}\ddot{\tilde{x}}(\tilde{t}) &= -\frac{\tau_{\rm{D}}}{\tau_{\rm{\Gamma}}} \int^{\tilde{t}}_0 \exp(-\tau_{\rm{D}}|\tilde{t}-\tilde{t}^{\prime}|/\tau_\Gamma)  \dot{\tilde{x}}(\tilde{t}^{\prime})d\tilde{t}^{\prime} \\
	& \qquad\qquad\quad\quad    -\frac{L}{k_{\text{B}}T}\frac{\partial}{\partial x}U(L\tilde{x}) +\tilde{F}_{\text{R}}(\tilde{t}^{\prime})	,
\end{split}
\end{equation}
where $\tilde{x}=x/L$, $\tilde{t}=t/\tau_{\rm{D}}$, and the dimensionless random force is $\tilde{F}_{\text{R}}(\tilde{t})= LF_{\text{R}}(\tau_{\text{D}}\tilde{t})/k_{\text{B}}T$. We recall that $\tau_{\rm{m}} = m/\gamma$ and $\tau_{\rm{D}} = \gamma L^2/k_{\rm{B}}T$. The autocorrelation of the dimensionless random force is 
\begin{equation}
\langle \tilde{F}_{\text{R}}(\tilde{t})\tilde{F}_{\text{R}}(\tilde{t}^{\prime})\rangle = \frac{\tau_D}{\tau_\Gamma}\exp(-\tau_{\rm{D}}|\tilde{t}-\tilde{t}^{\prime}|/\tau_\Gamma).
\end{equation}

\noindent For single-component exponential $\Gamma (t)$, Eq.~\ref{GLE} is equivalent to the following system of coupled equations:
\begin{equation}\label{ME}
\begin{split}
        m\ddot{x} &= -\nabla_xV(x, y),\\
        0  &=-\gamma\dot{y}+\nabla_yV(x, y)+\xi,
\end{split}
\end{equation}
where $V(x, y) = U(x) + \frac{\gamma}{2\tau_\Gamma }(y-x)^2$, and $U(x) = U_0[(x/L)^2-1]^2$ is the bi-stable potential well with barrier height $U_0$ and minimum-to-minimum distance $L$, which is the characteristic length used to rescale the position. $\xi $ is the Gaussian random force with $\langle \xi \rangle  = 0$ and $\langle \xi(t)\xi(t')\rangle =2k_{\rm{B}}T\gamma\delta(t-t')$. The rescaled, dimensionless form of Eq.~\ref{ME} corresponding to Eq.~\ref{GLER}, is
\begin{equation}\label{MER}
\begin{split}
        \frac{\tau_m}{\tau_D}\ddot{\Tilde{x}}(\Tilde{t}) &=- \frac{L}{k_{\text{B}}T}\frac{\partial}{\partial x}U(L\tilde{x})   +\frac{\tau_D}{\tau_\Gamma}[\Tilde{y}(\Tilde{t})-\Tilde{x}(\Tilde{t})] \\
    \dot{\Tilde{y}}(\Tilde{t}) &= -\frac{\tau_D}{\tau_\Gamma}\Tilde{y}(\Tilde{t})+\frac{\tau_D}{\tau_\Gamma}\Tilde{x}(\Tilde{t})+\Tilde{\xi}(\Tilde{t}),
\end{split}
\end{equation}
where $\langle \tilde{\xi}(t)\Tilde{\xi}(t')\rangle =2\delta(t-t')$. Eq.~\ref{MER} can be solved using the Runge-Kutta technique by introducing an auxiliary variable $z$ and modifying Eq.~\ref{MER} to include $\dot{\Tilde{x}}(\Tilde{t}) ={\Tilde{z}}(\Tilde{t})$.
\renewcommand{\theequation}{B\arabic{equation}} 
\setcounter{equation}{0}
\section{Relationship between $P^{\rm{ffp}}(\tau)$ and $P^{\rm{afp}}(\tau)$}
As introduced in the main article, for any first-to-first passage time $\tau^{\rm{ffp}}_n$, there exists a sequence of all-to-first passage events $\tau^{\rm{afp}}_{n,i}$ such that $\tau^{\rm{afp}}_{n,1} = \tau^{\rm{ffp}}_n$ and $\tau^{\rm{afp}}_{n,i} < \tau^{\rm{ffp}}_n$ for all $i>1$. Since any $\tau^{\rm{afp}}_{n,i}$ must associate with a $\tau^{\rm{ffp}}_n$ such that $\tau^{\rm{ffp}}_n \geq \tau^{\rm{afp}}_{n,i}$, the probability of observing an all-to-first $P_{\text{afp}}(\tau)$ is related to $P_{\text{ffp}}(\tau^{\prime})$ for all $\tau^{\prime} \geq \tau$:
\begin{equation}\label{PP_Rel_1}
P_{\text{afp}}(\tau) = C\int\limits_0^{\infty} P_{\text{ffp}}(\tau^{\prime}) {H}(\tau^{\prime} -\tau) d \tau^{\prime}.
\end{equation}
Here, $H(\tau^{\prime} -\tau)$ is the Heaviside function, defined such that 
\begin{equation}\label{It_G}
H(\tau^{\prime} -\tau) = \left\{ 
  \begin{array}{l l l l l}
  0, & \quad \tau^{\prime} -\tau < 0\\
  \quad \\
      1, & \quad \tau^{\prime} -\tau  \geq 0 \\
  \end{array}, \right.
\end{equation}
and $C$ is a normalization constant. We normalize such that 
\begin{equation}
\int\limits_{0}^{\infty}P_{\text{afp}}(\tau)d\tau = \int\limits_0^{\infty}C\int\limits_0^{\infty} P_{\text{ffp}}(\tau^{\prime}) {H}(\tau^{\prime} -\tau) d \tau^{\prime} d\tau = 1.
\end{equation}
Changing the order of integration leads to
\begin{equation}
C\int\limits_0^{\infty} P_{\text{ffp}}(\tau^{\prime}) \int\limits_0^{\infty} H(\tau^{\prime} -\tau) d\tau  d \tau^{\prime} = 1.
\end{equation}
To exploit the properties of the Heaviside function, we rewrite the inner integral such that
\begin{equation}
\begin{split}
\int\limits_0^{\infty} H(\tau^{\prime} -\tau) d\tau &= \int\limits_0^{\tau^{\prime}} H(\tau^{\prime} -\tau) d\tau +  \int\limits_{\tau^{\prime}}^{\infty} H(\tau^{\prime} -\tau) d\tau. \\
\end{split}
\end{equation}
The first term always satisfies $\tau \leq \tau^{\prime}$, where $H(\tau^{\prime} -\tau)=1$. Therefore, the integrand contributes a factor of $\tau^{\prime}$. The second term always satisfies $\tau > \tau^{\prime}$, where $H(\tau^{\prime} -\tau)=0$. The normalization constant is then
\begin{equation}\label{C_Norm}
C = \Bigg[ \int\limits_0^{\infty} P_{\text{ffp}}(\tau^{\prime}) \tau^{\prime}  d \tau^{\prime}\Bigg]^{-1}.
\end{equation}
By similarly exploiting the Heaviside function, we can rewrite Eq.\ref{PP_Rel_1} as
\begin{equation}
\begin{split}
P_{\text{afp}}(\tau) &= C\int\limits_0^{\tau} P_{\text{ffp}}(\tau^{\prime}) {H}(\tau^{\prime} -\tau) d \tau^{\prime} \\
& \quad + C\int\limits_{\tau}^{\infty} P_{\text{ffp}}(\tau^{\prime}) {H}(\tau^{\prime} -\tau) d \tau^{\prime}, \\
\end{split}
\end{equation}
where the first term is 0 since $\tau^{\prime} <\tau$. Therefore,
\begin{equation}
\begin{split}
P_{\text{afp}}(\tau) &=  C\int\limits_{\tau}^{\infty} P_{\text{ffp}}(\tau^{\prime}) d \tau^{\prime}. \\
\end{split}
\end{equation}
Including $C$ form Eq.\ref{C_Norm}, we acquire
\begin{equation}
\begin{split}
P_{\text{afp}}(\tau) &=  \frac{\int\limits_{\tau}^{\infty} P_{\text{ffp}}(\tau^{\prime}) d \tau^{\prime}}{ \int\limits_0^{\infty} P_{\text{ffp}}(\tau^{\prime}) \tau^{\prime}  d \tau^{\prime}}, \\
\end{split}
\end{equation}
which is Eq.\ref{Dist_Rel_main} from the main manuscript.



\clearpage
\newpage

\bibliography{Bib_File.bib}

\end{document}